\documentclass[aps,pra,twocolumn,floatfix,twoside]{revtex4}

\usepackage{graphicx}
\usepackage{dcolumn}
\usepackage{bm}

\begin{document}

\title{Dimers of ultracold two-component Fermi gases on magnetic-field
Feshbach resonance}
\author{Hongwei Xiong}
\affiliation{State Key Laboratory of Magnetic Resonance and Atomic and Molecular Physics,
Wuhan Institute of Physics and Mathematics, The Chinese Academy of Sciences,
Wuhan 430071, P. R. China}
\affiliation{Center for Cold Atom Physics, Chinese Academy of Sciences, Wuhan 430071,
China }
\affiliation{Graduate School of the Chinese Academy of Sciences}
\author{Shujuan Liu}
\affiliation{State Key Laboratory of Magnetic Resonance and Atomic and Molecular Physics,
Wuhan Institute of Physics and Mathematics, The Chinese Academy of Sciences,
Wuhan 430071, P. R. China}
\affiliation{Center for Cold Atom Physics, Chinese Academy of Sciences, Wuhan 430071,
China }
\author{Mingsheng Zhan}
\affiliation{State Key Laboratory of Magnetic Resonance and Atomic
and Molecular Physics, Wuhan Institute of Physics and Mathematics,
The Chinese Academy of Sciences, Wuhan 430071, P. R. China}
\affiliation{Center for Cold Atom Physics, Chinese Academy of
Sciences, Wuhan 430071, China }
\date{\today }

\begin{abstract}
At the location of a magnetic-field Feshbach resonance, a mixture gas of
fermionic atoms and dimers of fermionic atom pairs is investigated in the
unitarity limit where the absolute value of the scattering length is much
larger than the mean distance between atoms. The dynamic equilibrium of the
mixture gases is characterized by the minimum of the Gibbs free energy. For
the fermionic atoms and dimers with divergent scattering length, it is found
that the fraction of the dimers based on a very simple theory agrees with
the high fraction of zero-momentum molecules observed in a recent experiment
(M. W. Zwierlein et al, Phys. Rev. Lett. 92, 120403 (2004)). The dimeric gas
can be also used to interpret the frequency of the radial breathing mode
observed in the experiment by J. Kinast et al (Phys. Rev. Lett. 92, 150402
(2004)).


\end{abstract}

\maketitle

\section{introduction}

With the remarkable development of the cooling technique for two-component
atomic Fermi gases and magnetic-field Feshbach resonance \cite{THOMAS-FESH}
which can change both the strength and sign of the atomic interaction
characterized by the scattering length $a$ \cite{FESHBACH-THE}, experiments
have explored the crossover region between the strongly attractive and
repulsive interaction. For the spin mixture of the fermionic atoms, there
would be dimers of fermionic atom pairs in different form dependent on the
attractive or repulsive interaction between atoms. On the side of repulsive
interaction (BEC side), there exists molecule which is a short-range
fermionic atom pairs \cite{COLD-MOLECULES}. In several recent experiments
\cite{JOCHIM,JIN,MIT-mole}, the evidence for Bose-Einstein condensates of
diatomic molecules has been given. On the side of attractive interaction
(BCS side), one expects that the dimer is in a state of long-range fermionic
pairs analogously to the electronic Cooper pairs. The magnetic-field
Feshbach resonance gives us an important opportunity to investigate the
pairing phenomena in the ultracold Fermi gases especially the BCS-BEC
crossover which has been discussed in a lot of theoretical investigations
\cite{LEGGETT,NOZI,STOOF,TIMM,OHASHI1,MILSTEIN,STAJ,CARR,FALCO,WU} and
experimental researches \cite%
{JIN-fermion,MIT-fermion,GRIMM,SALOMON,THOMAS,GRIMM1,THOMAS-C2,CHENG,THOMAS-HEAT}
such as condensate fraction \cite{JIN-fermion,MIT-fermion}, collective
excitations \cite{THOMAS,GRIMM1,THOMAS-C2}, pairing gap \cite{CHENG} and the
recent measurement of heat capacity \cite{THOMAS-HEAT}.

In the BCS-BEC crossover, a particular interesting problem is the properties
of the ultracold gases at the location of the Feshbach resonance where the
scattering length between atoms is divergent. At or extremely close to the
location of the Feshbach resonance, the absolute value of the scattering
length $a$ can be comparable to or even much larger than the average
distance $\overline{l}$ between particles. In this situation, the ultracold
gas can be described in the unitarity limit \cite%
{CARLSON,HEISELBERG,HO1,HO,GHEM} where there is a universal behavior for the
system. The Feshbach resonant technique has given us the experimental means
to investigate the ultracold gas with $\left\vert a\right\vert >>\overline{l}
$. On resonance, the fraction of dimers in zero-momentum state is
investigated experimentally in \cite{JIN-fermion,MIT-fermion} after the
conversion of pairs of atoms into bound molecules. The experiment by MIT
group \cite{MIT-fermion} shows that there is a high fraction of dimers of
fermionic atoms extremely close to the Feshbach resonance. On resonance, the
frequency of the radial breathing mode is given in \cite{THOMAS} by
observing the collective oscillations of the system. In the present work,
based on a mixture of the atomic and dimeric gases in the unitarity limit $%
\left\vert a\right\vert >>\overline{l}$, a very simple theory is developed
to explain the recent experimental results in \cite{MIT-fermion} and \cite%
{THOMAS} at the location of the Feshbach resonance.

\section{chemical equilibrium on resonance}

In the present experiments on BCS-BEC crossover, an equal mixture of
fermionic atoms in two different spin states are confined in a harmonic trap
to realize atomic collisions. In thermal equilibrium, the system comprises
fermionic atoms and dimers, and its dynamic equilibrium is characterized by
the fact that the Gibbs free energy $G$ of the system is a minimum. Assuming
that $\mu _{F\uparrow }$ and $\mu _{F\downarrow }$ being the chemical
potential of the Fermi gases and $\mu _{D}$ being the chemical potential of
the dimeric gas, the minimum of the Gibbs free energy means that

\begin{equation}
2\mu _{F\uparrow }=2\mu _{F\downarrow }=\mu _{D}.  \label{chemical-potential}
\end{equation}%
The chemical potential of the dimeric gas is $\mu _{D}=\varepsilon _{D}+\mu
_{t}$ with $\varepsilon _{D}$ being the dimeric energy and $\mu _{t}$ being
the contribution to the chemical potential due to the thermal equilibrium of
the dimeric gas.

Assuming that $n_{F}$ is the total density distribution of the Fermi gases
without the presence of dimers and $x$ is the fraction of the dimers
immersed in the ultracold Fermi gases, the density distribution of the
dimeric gas $n_{D}$ and Fermi gas $n_{F\uparrow }$ (and $n_{F\downarrow }$)
in a spin state are respectively given by $n_{D}=xn_{F}/2$ and $n_{F\uparrow
}=n_{F\downarrow }=\left( 1-x\right) n_{F}/2$. For the unitarity limit $%
\left\vert a\right\vert >>\overline{l}$, there is a \textit{universality
hypothesis }(see \cite{HO})\textit{\ }that although the scattering length $a$
will play an important role, it will not appear in the final result of a
physical quantity such as chemical potential because it can be regarded as
infinity. In this case, the length scale $\overline{l}\sim n_{F}^{-1/3}$
rather than $a$ will appear in the final result of a physical quantity which
means a universal behavior for a system. By using the universality
hypothesis and local density approximation, at zero temperature, the
chemical potential $\mu _{F\uparrow }$ ($=\mu _{F\downarrow }$) of the Fermi
gas is given by

\begin{equation}
\mu _{F\uparrow }=\left( 1+\beta _{1}\right) \frac{\hbar ^{2}\left( 6\pi
^{2}\right) ^{2/3}}{2m}\left( \frac{\left( 1-x\right) n_{F}}{2}\right)
^{2/3}+V_{ext}\left( \mathbf{r}\right) ,  \label{chemical-fermi}
\end{equation}%
where $V_{ext}\left( \mathbf{r}\right) $ is the external potential of the
fermionic atoms. Without $\beta _{1}$, the above expression gives the
chemical potential of an ideal Fermi gas in the local density approximation.
The parameter $\beta _{1}$ shows the role of the extremely large scattering
length $a$ in the unitarity limit. The parameter $\beta _{1}$ was firstly
measured in \cite{THOMAS-FESH} with a careful experimental investigation of
the strongly interacting degenerate two-component Fermi gases near the
Feshbach resonance. $\beta _{1}$ has been also calculated with different
theoretical methods \cite{CARLSON,HEISELBERG}.

The scattering length $a_{D}$ ($\approx 0.6a$) \cite{PETROV} of the dimers
is comparable to the scattering length $a$ between fermionic atoms, thus the
dimeric gas will also show a universal behavior on resonance. In the
unitarity limit, for the bosonic dimer gas, the scattering length $a_{D}$
should not appear in the final result of a physical quantity too. According
to the\ universality hypothesis, we assume here that the bosonic dimer gas
shows an analogous universal behavior with the Fermi gas (see also \cite{HO}%
). Based on this assumption, one has the chemical potential for the dimeric
gas at zero temperature

\begin{equation}
\mu _{D}=\left( 1+\beta _{2}\right) \frac{\hbar ^{2}\left( 6\pi ^{2}\right)
^{2/3}}{2\times 2m}\left( \frac{xn_{F}}{2}\right) ^{2/3}+2V_{ext}\left(
\mathbf{r}\right) +\varepsilon _{D}.  \label{chemical-dimer}
\end{equation}%
The parameter $\beta _{2}$ in $\mu _{D}$ is also due to the large scattering
length $a_{D}$ between dimers and it can be calculated in the unitarity
limit. Near the Feshbach resonance, the dimeric energy $\varepsilon _{D}\sim
\hbar ^{2}/ma^{2}$. Thus, on resonance, $\varepsilon _{D}$ can be omitted in
the above expression. From the dynamic equilibrium condition $2\mu
_{F\uparrow }=\mu _{D}$, one gets the following general equation to
determine the fraction of dimers at zero temperature and Feshbach resonance:

\begin{equation}
4\beta _{r}\left( 1-x\right) ^{2/3}=x^{2/3},  \label{fraction}
\end{equation}%
where $\beta _{r}=\left( 1+\beta _{1}\right) /\left( 1+\beta _{2}\right) $.
We show the relation between the fraction of dimers $x$ and different value
of $\beta _{r}$ in Fig. 1.

In the presence of the dimers, the chemical potential of the system is
obviously lower than the case without the dimers. Thus, the mixture gas of
the atoms and dimers is a stable state of the system. For $\left\vert
a\right\vert >>\overline{l}$, there are three important characteristics for
the dimeric gas:

(i) \textit{The above result for the fraction of the dimeric gas is a
general result once the system is in the unitarity limit.}

(ii) \textit{There is a quite high fraction of the dimers as shown in Fig.1.}
In the present experiments, the fraction of dimers is determined by the
molecules in zero-momentum state after a fast magnetic field transfer (the
magnetic field is swept below the Feshbach resonant magnetic field $B_{0}$
so that the scattering length between fermionic atoms becomes positive) to
create bound molecules from the dimers. After the dimer-molecule conversion,
in \cite{MIT-fermion} the maximum fraction of the molecules in zero-momentum
state is observed to be $80\%$ which agrees with our theoretical result.
From this experimental result, $\beta _{r}$ is estimated to be $0.63$.

(iii) \textit{The dimer comprising two fermionic atoms is quite stable.} At
zero temperature, the dimeric gas is immersed in the degenerate Fermi gases.
Due to Pauli blocking comes from the degenerate Fermi gases, any dimer can
not be dissociated into two fermionic atoms once the equilibrium is attained
so that $2\mu _{F\uparrow }=\mu _{D}$. In fact, the stability of the dimeric
gas is consistent with the experiment \cite{MIT-fermion} that extremely
close to the Feshbach resonance there is no obvious decreasing of the
molecules in zero-momentum state after the dimer-molecule conversion process
even after $10$ \textrm{s }hold time of the final magnetic field. Due to the
stability and Pauli blocking, the dimer-molecule conversion will always
convert the fermionic atom pairs in a same dimer into bound molecules. Thus
at zero temperature, all the dimers will be converted into molecules with
zero-momentum state during the dimer-molecule conversion process. This means
that the fraction of dimers in thermal equilibrium investigated here can be
used to explain the experimental result of the fraction of zero-momentum
molecules after the dimer-molecule conversion.

\begin{figure}[tbp]
\includegraphics[width=0.8\linewidth,angle=270]{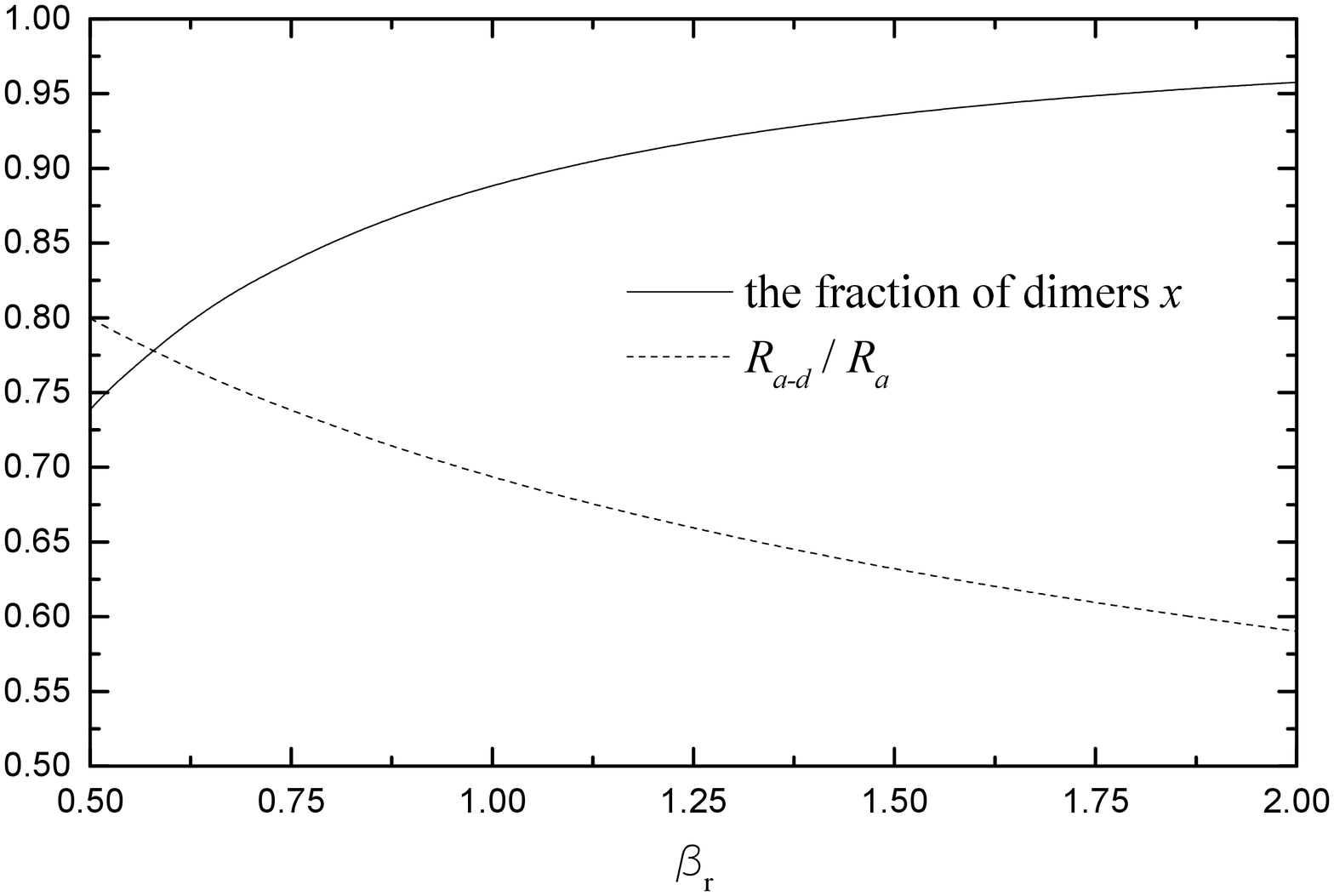}
\caption{The solid line shows the fraction of the dimers at the
location of the Feshbach resonance and zero temperature with
different parameter $\beta _{r}$ which can be calculated in the
unitarity limit. The high fraction of the dimers agrees with the
experiment \cite{MIT-fermion} where the maximum fraction of the
molecules in zero-momentum state is observed to be $80\%$ after
the dimers are converted into molecules. The dashed line shows the
effect of the existence of the dimeric gas on the width of gases
confined in the optical trap. We see that the existence of dimers
has an obvious effect of decreasing the width of the gases in the
trap.}
\end{figure}

In a recent experiment \cite{THOMAS}, the frequency of a radial breathing
mode is observed extremely close to the Feshbach resonance and found to
agree well with the theoretical predication based on hydrodynamic theory in
the unitarity limit \cite{STRINGARI}. Combining with the experimental result
in \cite{MIT-fermion} (also shown in the present work) that there is a high
fraction of molecules in zero-momentum state after the dimer-molecule
conversion process, we see that the dimeric gas should play a dominant role
in determining the frequency of collective oscillations due to its high
fraction. For the radial breathing mode, the agreement of the experiment
with theoretical result in the unitarity limit shows that it is reasonable
to describe the bosonic dimer gas by the unitarity hypothesis which is used
to give Eq. (\ref{chemical-dimer}). In fact, an analogous form of Eq. (\ref%
{chemical-dimer}) is used to calculate the frequency of the radial breathing
mode in \cite{STRINGARI}, whose result agrees well with the experiment in
\cite{THOMAS}. Different from the theoretical model in \cite{STRINGARI},
however, in the present work on resonance, the dimeric gas should play a
dominant role in the frequency of the collective oscillations at zero
temperature, rather than the Fermi gases.

On resonance, the mixture of the fermionic atoms and dimers will play a role
in the width of the gases confined in the harmonic trap. From the dynamic
equilibrium $2\mu _{F\uparrow }=\mu _{D}$, the width of the Fermi gas is
equal to that of the dimeric gas. Assume that $R_{a-d}$ is the width of the
mixture gases for the atoms and dimers, while $R_{a}$ is the width of the
gases without considering the existence of the dimers. After a simple
calculation, one has the ratio $R_{a-d}/R_{a}=\left( 1-x\right) ^{1/6}$. The
dashed line in Fig.1 shows $R_{a-d}/R_{a}$ as a function of $\beta _{r}$. We
see that the existence of dimers in dynamic equilibrium will decrease the
width of the gases.

\section{summary}

In summary, the role of the dimers of fermionic atoms is discussed extremely
close to the Feshbach resonance where both the fermionic atoms and dimers
show a universal behavior based on the universality hypothesis. The dynamic
equilibrium of the mixed atom-dimer gas is considered for the case of a
harmonic trap through the analysis of the Gibbs free energy. Based on this
simple theoretical model, it is found that the high fraction of the stable
dimeric gas in the present work agrees well with the experiment \cite%
{MIT-fermion} where there is a high fraction of molecules in zero-momentum
state after the conversion of the dimers into molecules. Extremely close to
the Feshbach resonance, the dimeric gas plays a dominant role for the
frequency of the radial breathing mode observed in \cite{THOMAS}. On
resonance, the stability of the dimeric gas would be useful in the creation
of molecular BEC on the BEC side and condensate of atomic Cooper pairs on
the BCS side. Crossing the Feshbach resonant magnetic field $B_{0}$ slowly
from above (or below) would contribute to the creation of molecular BEC (or
condensate of atomic Cooper pairs) because there is a formation of a stable
dimeric gas when $B_{0}$ is swept cross slowly. For example, in \cite{JIN},
the molecular BEC is created after the magnetic field is swept slowly from
above to below $B_{0}$.

\begin{acknowledgments}
This work is supported by NSFC under Grant No. 10205011 and NBRPC
under Grant No. 2001CB309309.
\end{acknowledgments}

\end{document}